\documentclass[electronic]{vgtc}             %

\ifpdf%
  \pdfoutput=1\relax                   %
  \usepackage{graphicx}                %
  \DeclareGraphicsExtensions{.pdf,.png,.jpg,.jpeg} %
\else%
  \usepackage{graphicx}                %
  \DeclareGraphicsExtensions{.eps}     %
\fi%

\graphicspath{{figures/}{pictures/}{images/}{./}} %

\usepackage{microtype}                 %
\PassOptionsToPackage{warn}{textcomp}  %
\usepackage{textcomp}                  %
\usepackage{mathptmx}                  %
\usepackage{times}                     %
\usepackage{cite}                      %
\usepackage{tabu}                      %
\usepackage{booktabs}                  %
\usepackage{enumitem}
\usepackage{xspace}
\usepackage{wrapfig}

\onlineid{0}

\vgtccategory{Research}

\vgtcinsertpkg
\definecolor{linkColor}{RGB}{6,125,233}

\title{Evaluating Cardiovascular Surgical Planning \\in Mobile Augmented Reality }

\newcommand{\authorgap}{\hspace{10pt}}
\renewcommand\footnotemark{}

\author{
    Haoyang Yang\textsuperscript{\textrm 1} %
    \thanks{\textsuperscript{\textrm 1}Georgia Institute of Technology. \{\href{mailto:alexanderyang@gatech.edu}{alexanderyang}$\,\mid\, $\href{mailto:pratham@gatech.edu}{pratham}$\,\mid\, $\href{mailto:jpleo122@gatech.edu}{jpleo122}$\,\mid$
    \newline \textcolor{white}{.} \hspace{11pt}
    \href{mailto:zzhou406@gatech.edu}{zzhou406}\hspace{-2pt}
    $\,\mid\, $\href{mailto:mdass3@gatech.edu}{mdass3}\hspace{-2pt}
    $\,\mid\, $\href{mailto:aupadhayay3@gatech.edu}{aupadhayay3}\hspace{-2pt}
    $\,\mid\, $\href{mailto:polo@gatech.edu}{polo}\}@gatech.edu}
    \authorgap
    Pratham Darrpan Mehta\textsuperscript{\textrm 1} \authorgap
    Jonathan Leo\textsuperscript{\textrm 1} \authorgap
    Zhiyan Zhou\textsuperscript{\textrm 1} \authorgap
    Megan Dass\textsuperscript{\textrm 1} \authorgap\\
    Anish Upadhayay\textsuperscript{\textrm 1} \authorgap
    Timothy C. Slesnick\textsuperscript{\textrm 2} %
    \thanks{\textsuperscript{\textrm 2}Children's Healthcare of Atlanta. \href{mailto:SlesnickT@kidsheart.com}{SlesnickT@kidsheart.com},
    \newline \textcolor{white}{.} \hspace{11pt}
    \href{mailto:Fawwaz.Shaw@choa.org}{Fawwaz.Shaw@choa.org}}  \authorgap
    Fawwaz Shaw\textsuperscript{\textrm 2} \authorgap
    Amanda Randles\textsuperscript{\textrm 3} %
    \thanks{\textsuperscript{\textrm 3}Duke University. \href{mailto:amanda.randles@duke.edu}{amanda.randles@duke.edu}} \authorgap
    Duen Horng Chau\textsuperscript{\textrm 1} \authorgap
}

\newcommand{\tool}[0]{\textsc{CardiacAR}\xspace{}}

\abstract{
Advanced surgical procedures for \textit{congenital heart diseases} (CHDs) require precise planning before the surgeries. The conventional approach utilizes 3D-printing and cutting physical heart models, which is a time and resource intensive process. While rapid advances in augmented reality (AR) technologies have the potential to streamline surgical planning, there is limited research that evaluates such AR approaches with medical experts. This paper presents an evaluation with 6 experts, 4 cardiothoracic surgeons and 2 cardiologists, from
\textit{Children's Healthcare of Atlanta} (CHOA)\textit{ 
Heart Center} to validate the usability and technical innovations of \tool{}, a prototype mobile AR surgical planning application. Potential future improvements based on user feedback are also proposed to further improve the design of \tool{} and broaden its access.

} %

\CCScatlist{
  \CCScatTwelve{Human-centered computing}{Human computer interaction (HCI)}{Interaction paradigms}{Augmented reality}
}

\begin{document}

\firstsection{Introduction}

\maketitle
Surgical procedures for \textit{congenital heart diseases} (CHDs) require the surgeons to have a deep understanding of the complex cardiac anatomy, necessitating the process of surgical planning  \cite{Bartel2018MedicalTP}. 
A typical planning process utilizes a 3D-printed heart model to visualize the morphological features of the patient’s heart\cite{riggs_3d-printed_2018}. However, producing physical models can be time and resource intensive \cite{Kappanayil2017ThreedimensionalprintedCP,yoo_3d_2021}. 
While interactive surgical planning tools based on extended
reality (XR) technologies, including virtual reality (VR), mixed reality (MR), and augmented reality (AR), have the potential to expedite planning, there is limited work that evaluates such tools with feedback directly from surgeons and cardiologists \cite{Leo2021InteractiveCS}. 
To fill this research gap, we have conducted a usability evaluation with 6 medical experts for 
\tool{}\cite{Leo2021InteractiveCS} (\autoref{fig:teaser}), an iOS application that enables interactive surgical planning on mobile devices through AR. Built on top of the iOS mobile ARKit platform, \tool{} offers a suite of interactive tools, such as an enhanced AR view of patients’ 3D heart models in real-life environments (\autoref{fig:teaser}A), real-time omni-directional slicing of models (\autoref{fig:teaser}B), and virtual annotation (\autoref{fig:teaser}C), to assist surgical planning.
\tool{} is  open-source and publicly available at \textcolor{linkColor}{\url{https://github.com/poloclub/CardiacAR}}.
Our contributions:

\begin{figure}[t]
  \includegraphics[width=\linewidth]{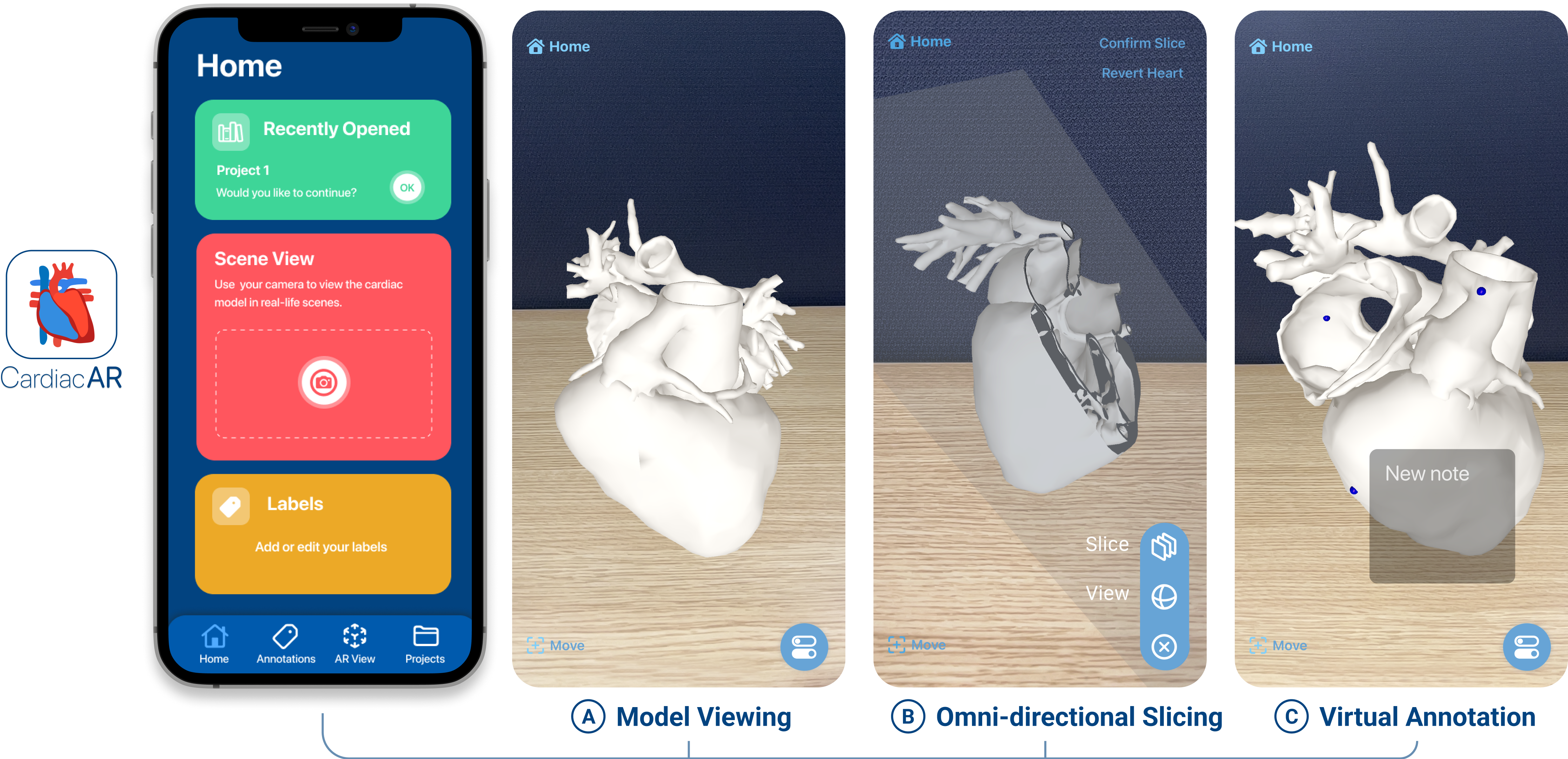}
  \caption{\tool{}, an iOS augmented reality application that enables users to perform interactive surgical planning on mobile devices.  %
  }
  \label{fig:teaser}
\end{figure}

\begin{itemize}[topsep=1pt, itemsep=0mm, parsep=3pt, leftmargin=9pt]

   \item 
     \textbf{First evaluation of mobile AR surgical planning tool with medical experts.} 
     To the best of our knowledge, this research is the first to evaluate the usability of mobile AR technologies  with medical experts. We recruited four cardiothoracic surgeons and two cardiologists from
\textit{Children's Healthcare of Atlanta} (CHOA)\textit{ 
Heart Center}.
     Their positive feedback highlighted mobile AR's strong potential to support cardiovascular surgical planning
     and benefit real surgery scenarios.

    \item 
    \textbf{Technical discoveries enabling mobile AR for surgical planning.} 
    Through developing \tool{} and planning for its evaluation with mobile devices that surgeons commonly use (e.g., iPhones, iPads), we made two important technical discoveries that enabled the study:
    
    \begin{enumerate}[topsep=1pt, itemsep=0mm, parsep=3pt, leftmargin=9pt]
    \item
   \textbf{Innovative real-time omni-directional slicing functionality.} Conventional method of uni-directionally slicing the 3D physical model is only able to produce a few cross-sectional views per physical artifact\cite{Sun2019PersonalizedTP}.
    In contrast, \tool{} supports \textbf{preview slicing}, which provides real-time visualization of the slicing plane's orientation and highlights the planar cross section with surface shaders.
    Model slicing is not natively supported within  ARKit
or SceneKit.
While planar slicing for simpler solids using software like Unity is possible, the approach is not compatible with 3D heart models, as they contain hollow chambers and tubes. 
We experimented with approaches such as Boolean operations for 3D solids and voxel arrays 
but they are too slow (e.g., seconds per operation) 
and do not support hollow geometries like the 3D hearts (forcibly applying such operations on a heart model often led to anomalous surfaces in the resulting model geometry). 
Thus, we developed our novel approach that enabled real-time omni-directional slicing (\autoref{fig:teaser}B) using surface shader to paint the cross section in a non-reflective dark color, highlighting the inner surfaces of the heart model's geometry and achieving the correct visual effects.
    \item
    \textbf{Streamlining deployment process and increasing accessibility of the application.}
    \tool{} is developed natively in XCode and is open-sourced under the MIT license. The ubiquity of iOS mobile AR platform allows us to design interaction gestures that are familiar and easy to use.
\tool{} is released on TestFlight, a widely-available platform 
that
enables easy distribution and testing over the iterative development process. Through collaborating with medical experts, we have discovered that TestFlight's asynchronous testing capabilities greatly facilitate the testing and feedback collection process, essential when working with doctors with busy  schedules.

   \end{enumerate}
\end{itemize}

\section{Usability Evaluation with Medical Experts}
In collaboration with \textit{Children's Healthcare of Atlanta} (CHOA), 
we recruited  six medical experts who had extensive knowledge in cardiovascular surgeries to evaluate the usability of \tool{}---
four were cardiovascular surgeons, 
and two were cardiologists. 
CHOA provided the 3D heart model used in the study; the model was constructed by CHOA  using a patient's de-identified medical imaging data. 
The study was approved by Georgia Tech’s IRB, with data collection in accordance with the official institute policies.

\subsection{Procedure}

We conducted the user study in person in a quiet room at CHOA, the participants' work premises. 
The study was conducted across two group sessions. The first session was with the four cardiovascular surgeons, 
and the second one with the two cardiologists.
We originally planned to conduct the study with each participant individually but we had adapted our study design to a group-based one due to their limited availability.
Both sessions were recorded using a video camera, and the devices used were also screen recorded.
Before commencing the study, each participant was asked to review and sign a consent form.
We gave each participant an iOS device (iPhone or iPad) with the latest iOS version and \tool{} installed. 
Then, we provided a brief tutorial of the \tool{}'s four features, which they would then try in sequence:
(1) \textit{importing model}: import a 3D heart model into \tool{}, and reposition it using the ``Move'' button;
(2) \textit{model viewing}: rotate and resize the model using finger gestures;
(3) \textit{model slicing}: switch to slicing mode, rotate and translate the slicing plane to preview possible slicing angles, and then confirm a slice;
(4) \textit{virtual annotation}: tap on the model to create a note.
Lastly, the participants had an additional five minutes to freely explore the application.

\subsection{Results and Key Findings} 
For each feature, 
we asked the participants to rate its usability and usefulness using a 5-point Likert scale (5 being ``the best'').
We also asked the participants to elaborate on how the feature may help with cardiovascular surgery use cases and how they may want the feature improved.
We summarize the participants' feedback and our discoveries into the following main categories.

\smallskip
\noindent
\textbf{Omni-directional slicing makes visualization of cardiovascular anatomy easier.} 
Using \textit{preview slicing}, participants were able to visualize specific cross-sections and angles of the heart. 
Both participant groups---surgeons and cardiologists---strongly agreed that this feature would benefit real-life scenarios of surgical planning, such as helping them gain a better understanding of the geometry of the heart than the conventional approach using 3D-printed physical heart models. 
They gave an average Likert rating of 4.5 for usability across the two groups.  
They found it helpful that \tool{} allowed them to easily preview possible slicing angles and translations before confirming the slice, enabling them to find the best perspective to perform such a slice; one cardiologist commented that they could then pass the sliced model to their colleagues for further inspection and additional slicing.
The surgeons suggested that the gesture sensitivity should be lowered as they initially found it to be high, which resulted in rapid movements of the model.

\smallskip
\noindent
\textbf{Mobile nature of tool facilitates portability.} 
The participants appreciated that \tool{} was designed for mobile devices.
With the portability provided by the iPhone and iPad, 
the participants were comfortable moving the device across the physical space to ``enter'' the model and obtain an ``inside'' view of the heart model while 
viewing and manipulating the model.
Both participant groups rated the ease of use of the \textit{model viewing} feature positively with rating of 4.
All participants rated the ease of use of \tool{}'s annotation feature highly, with a rating of 5.
The cardiologists commented that this feature would be very helpful in practical scenarios for labelling and demarcating important regions. 
Both groups suggested that the annotation feature could be further developed and offered in its own dedicated ``mode'' (e.g., similar to the \textit{slicing} and \textit{viewing} modes),
so to help separate the annotation interactions from those of the other features.  
Finally,
all participants commented that \tool{}'s portability and annotation feature would make it a convenient tool to use in educational settings.

\smallskip 
\noindent
\textbf{Easy model import supports patient-specific analysis.} 
The model import feature was very well received across all participants, with an average Likert rating of 4.5 for usability across both groups. 
The participants believed that this feature would allow them to explore different heart models from a variety of patients in an efficient manner. 
This would facilitate patient-specific heart diagnosis and allow surgeons and cardiologists to effectively examine anomalies present in the heart. 
The cardiologists added that the ability for \tool{} to directly import models from 
HIPAA--compliant\footnote{Health Insurance Portability and Accountability Act} cloud storage solutions, such as Microsoft OneDrive,  
facilitate collaboration with other medical professionals situated in different locations.

\smallskip
\noindent
\textbf{Participants' technical needs inform different levels of AR immersion.} 
An interesting observation was the difference in AR-related requirements across both groups. 
Throughout the study, the surgeons preferred that the physical world ``background'' (captured by the camera) behind the model not be visible so as to maintain focus on the model. 
However, the cardiologists had a different opinion  in that they believed that viewing the background helped them better anchor the model in the physical space, so they can more easily examine and interact with it.

\section{Conclusion and Ongoing Work}
The paper presents a follow-up usability evaluation of \tool{ }with medical experts. The user feedback from the evaluation demonstrates the efficacy of \tool{} in helping users improve spatial understanding of the model while making interactive planning more convenient and efficient.
Based based on feedback from the medical experts, we have proposed several improvements to the project. Certain gesture movements and user interface for the slicing plane and panning feature can be refined. A larger-scale evaluation with more participants and types of surgeries can also help assess its usability in different scenarios.

\bibliographystyle{abbrv-doi}
\bibliography{CardiacAR}
\end{document}